\documentstyle[11pt,paspconf]{article}

\begin{document}
\newcommand{\chvcs}{{C IV}-HVCs}
\newcommand{\degr}{$^{\circ}$}
\newcommand{\et}{{\it et al.~}}
\newcommand{\hhvcs}{{H I}-HVCs}
\newcommand{\kms}{\,km\,s$^{-1}$}     
\newcommand{\lam}{$\lambda$}
\newcommand{\lya}{Ly$\alpha$\ }
\newcommand{\subsun}{\mbox{$_{\odot}$}}
\newcommand{\twid}{\,$\sim$\,}
\newcommand{\vlsr}{V$_{\sc LSR}$}
\newcommand{\phn}{\phantom{0}}

\hyphenation{Origins}
\hyphenation{Hopkins}

\def\hi{H\,{\sc i}}
\def\hii{H\,{\sc ii}}
\def\di{D\,{\sc i}}
\def\ci{C\,{\sc i}}
\def\cii{C\,{\sc ii}}
\def\ciii{C\,{\sc iii}}
\def\ni{N\,{\sc i}}
\def\nii{N\,{\sc ii}}
\def\niii{N\,{\sc iii}}
\def\oi{O\,{\sc i}}
\def\ovi{O\,{\sc vi}}
\def\alii{Al\,{\sc ii}}
\def\arii{Ar\,{\sc ii}}
\def\ari{Ar\,{\sc i}}
\def\mgii{Mg\,{\sc ii}}
\def\pii{P\,{\sc ii}}
\def\piii{P\,{\sc iii}}
\def\piv{P\,{\sc iv}}
\def\pv{P\,{\sc v}}
\def\cli{Cl\,{\sc i}}
\def\clii{Cl\,{\sc ii}}
\def\siii{S\,{\sc iii}}
\def\siv{S\,{\sc iv}}
\def\svi{S\,{\sc vi}}
\def\arii{Ar\,{\sc ii}}
\def\criii{Cr\,{\sc iii}}
\def\feii{Fe\,{\sc ii}}
\def\feiii{Fe\,{\sc iii}}

\title{The Far Ultraviolet Spectroscopic Explorer: Mission Overview
and Prospects for Studies of the Interstellar Medium and High 
Velocity Clouds}

\author{Kenneth R. Sembach}
\affil{Department of Physics \& Astronomy, The Johns Hopkins University,
Baltimore, MD 21218, U.S.A}

\begin{abstract}
The {\it Far Ultraviolet Spectroscopic Explorer} (FUSE) is a NASA astronomy
mission that will explore the 905--1187\AA\ wavelength region 
at high spectral 
resolution.  Funded by NASA's Explorer Program, this {\it Origins} mission 
is scheduled for a 1999 launch and at least three years of operations.  The
development of FUSE is being led by the Johns Hopkins University, with major
contributions to the program from the University of Colorado,
the University of California-Berkeley, the space agencies of Canada and 
France, and corporate partners.
\smallskip
\\
FUSE will have approximately 10,000 times the sensitivity of its
pioneering predecessor, {\it Copernicus}, which operated in the 1970s.  
Much of
the FUSE Science Team observing time will be dedicated to studying the 
interstellar medium of the Milky Way and Magellanic Clouds.  Observations of 
high velocity clouds play an important role in the FUSE program.  In this 
paper, I outline some of the FUSE Science Team plans for observing 
HVCs.  Simple absorption line models are also provided 
for investigators seeking
to identify atomic and molecular species in this wavelength region.

\end{abstract}

\keywords{instrumentation: spectrographs --- ISM: abundances --- ISM: clouds
--- ISM: atoms --- ISM: molecules --- Galaxy: evolution --- Galaxy: halo 
--- ultraviolet: ISM}

\section {Introduction}
With the launch of the {\it Far Ultraviolet Spectroscopic Explorer}
(FUSE) in 1999,
the astronomical community will have access to the first long duration
satellite devoted to high-resolution 
($\lambda/\Delta\lambda$\,$\approx$\,24,000--30,000) spectroscopic studies 
of the far ultraviolet (far-UV) universe since the {\it Copernicus} mission.  
FUSE will have a point source sensitivity approximately 10,000 times that of 
{\it Copernicus}, which will allow systematic studies of distant regions in 
the Milky Way and other galaxies to be conducted at high spectral 
resolution in 
the 905--1187 \AA\ bandpass for the first time.  FUSE will also be able to 
observe the far-UV light from distant quasars and active galactic nuclei. 

The far-UV wavelength region is rich in spectral line diagnostics of plasmas 
ranging in temperature from 10$^1$--10$^6$~K.  It encompasses the Lyman
series of \hi\ and \di, as well as resonance lines of the heavy element 
species \ci, \cii, \ciii,
\ni, \nii, \niii, \oi, \ovi, F\,{\sc i}, \mgii, \alii, Si\,{\sc ii}, 
\piii, \piv, \pv, \siii, 
\siv, \svi, \cli, \clii, \ari, \arii, \feii, and \feiii.  
Weak lines of other heavy elements (e.g., Cr, Mn, Ni) and low excitation
fine-structure lines of \ci, \cii, \ni, and \nii\ may be visible along 
some sight lines.
The bandpass also contains 
molecular lines in the Lyman and Werner systems of H$_2$ and HD,
and the $B^1\Sigma^+_u-X^1\Sigma^+_g$ and $C^1\Sigma^+_u-X^1\Sigma^+_g$ bands 
of CO.

\section{FUSE Team Science Investigations} 
During the first three years of operation, approximately half of the FUSE
observing time will be used by Guest
Investigators selected through a competitive peer-reviewed process, 
which at the 
time of writing has been completed for the first observing cycle.
A small amount (roughly 10\%) of the total observing time 
has also been reserved for the French and Canadian astronomical 
communities.  The remainder will be used by the FUSE Science
Team, which will undertake
several large science investigations as well as a number of 
moderate-sized programs designed to study specific astronomical objects
or phenomena.  The two primary science programs include:

\smallskip
\noindent 1) A study of the D/H ratio and its dependence
upon the chemical evolution of the interstellar gas in the Milky Way 
and intergalactic gas in the low redshift universe.

\smallskip
\noindent 2) A study of the origins and properties of hot 
(T\,$\sim$\,10$^5$--10$^6$~K) interstellar gas in the Milky Way and 
Magellanic Clouds as traced through \ovi\ absorption and emission.

\vspace{8.7cm}
\begin{figure}[!h]
\includegraphics{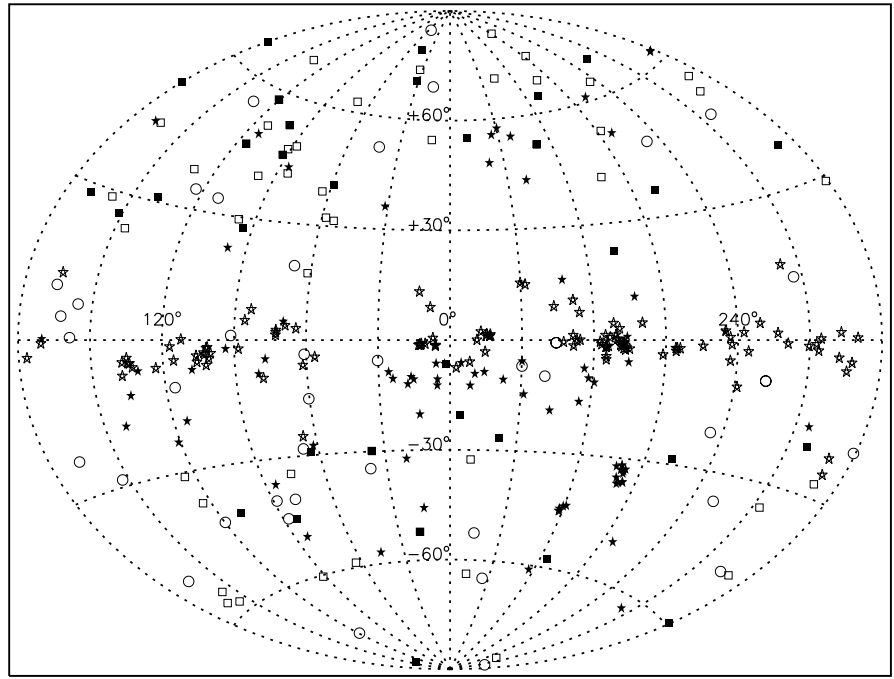}
\caption{\footnotesize Aitoff projection of the FUSE Team \ovi\ and D/H 
program sight lines
in Galactic coordinates. The Galactic center is at the center of the figure.  
{\it Open circles}: Local ISM (d\,$<$\,300 pc); {\it Open stars}: 
Galactic disk ($|$z$|$\,$<$\,300 pc); {\it Filled stars}: 
Galactic halo ($|$z$|$\,$>$\,300 pc); {\it Filled squares}: \ovi\ and D/H
extragalactic; {\it Open squares}: D/H extragalactic snapshots.}
\end{figure}

To conduct these two comprehensive studies, the FUSE Science Team will observe 
a large 
number of objects in the Galactic disk and halo.  
Figure~1 is a projection of 
these sight lines onto the sky.  The sight lines sample 
the interstellar medium (ISM) in a variety of directions over distances 
ranging from 
a few parsecs to tens of kiloparsecs.  The types of regions to be explored 
include translucent molecular cloud envelopes, cool neutral clouds, warm 
neutral clouds, the ionized ISM of the Galactic disk and halo, 
supernova remnants, and hot gas associated with bubbles/supershells in the 
Magellanic Clouds. The FUSE Team will also observe sight lines to 
quasars and active galactic nuclei to study high velocity clouds,
the distant Galactic halo and
intergalactic gas in the low redshift universe.  A portion of the observing 
time will be used for short snapshot exposures to determine the best 
sight lines for extended observations.
The primary objectives of the D/H and \ovi\ programs are listed below in
\S2.1 and \S2.2.

\subsection{Goals of the FUSE Team D/H Program}
Most of the deuterium in the Universe was created within a few minutes of the 
Big Bang, and it is generally believed that the net abundance of deuterium 
decreases with time due to stellar processing (astration). The
present day value of the deuterium abundance should therefore reflect the 
imprint of Big Bang nucleosynthesis as
well as the subsequent chemical evolution of the Universe. To understand
this history, it is necessary to disentangle local effects from global 
effects on the D/H ratio, and to integrate the results for a large number 
of regions into a coherent description of the zero-redshift abundance
of deuterium.  To this end, the FUSE Team has outlined the following 
objectives:

\smallskip \noindent
1) Quantify the effects of local environmental conditions and
processes (e.g., astration, fractionation, ionization, metal
production) on the measured abundance of deuterium.

\smallskip \noindent
2) Determine whether the D/H ratio varies within the Milky
Way and the implications variability would have for the chemical
evolution and mixing of the interstellar medium.

\smallskip \noindent
3) Determine the D/H ratio in environments with a range of metallicities
to use as zero redshift benchmarks for D/H values obtained for low
metallicity systems at high redshift.

\smallskip \noindent
4) Integrate the Milky Way D/H results with chemical evolution models
to provide a clearer understanding of galactic chemical evolution, the 
baryonic content of the Universe, and Big Bang nucleosynthesis.

\subsection{Goals of the FUSE Team \ovi\ Program}

\ovi\ is the primary far-UV line diagnostic of hot 
(T\,$\sim$\,3$\times$10$^5$ K),
collisionally ionized gas in the interstellar medium.  The production,
distribution, and quantity of interstellar gas in this temperature
regime outside the local region of the ISM surveyed by
{\it Copernicus} is unknown.  Since the processes that create hot gas
(e.g., supernovae) are closely related to the physical properties
of the ISM, star-formation, heavy element production, the transport of 
mass and energy, and the chemical evolution of galaxies, the FUSE Team has 
defined the following objectives:

\smallskip \noindent
1) Study the physical processes that create interstellar \ovi\
and quantify  the role of the hot ISM in controlling the physical 
properties, distribution, and chemical evolution of gas in the Galaxy.

\smallskip \noindent
2) Study the transport of energy and matter in the Galaxy and the 
effects of a ``disk-halo connection'' on the maintenance of a
hot Galactic corona.

\smallskip \noindent
3) Determine the three dimensional distribution of local hot gas and
study the hot/warm gas interfaces at the Local Cloud/Local Bubble 
boundaries.

\smallskip \noindent
4) Understand how hot interstellar gases in the Milky Way, LMC, and SMC
are related to large scale ISM structures (supernova remnants, 
supershells, radio loops, etc.), and apply this knowledge to studies of 
galaxies and quasar absorption line systems.

\subsection{Focused Investigations and the FUSE Archive}
In addition to the D/H and \ovi\ programs, the FUSE Science Team will
conduct studies in a number of areas using data from the key
programs as well as supplemental observations.  These investigations will
include studies of 
H$_2$ and the CO/H$_2$ ratio, hot star winds and atmospheres,
cool star chromospheres, supernova remnants (including SN 1987A), 
cooling flows, active galactic nuclei, jets and 
circumstellar disks, and planetary atmospheres. A high-resolution measurement 
of the He\,{\sc ii} Gunn-Peterson effect will also be made 
if in-orbit background levels are sufficiently low.

For the first cycle of operations, 63 Guest Investigator programs have
been selected to address many intriguing astronomical questions.  
The data from Team and GI investigations 
will provide a wealth of information long after 
the mission has ended.  FUSE data will have a six month proprietary period 
and will be archived at the Space Telescope Science Institute in
Baltimore, Maryland.

\section{FUSE and High Velocity Clouds}

The properties of high velocity clouds (HVCs) are poorly known despite several 
decades of study.  This is due in large part to the general lack of 
spectroscopic information at ultraviolet wavelengths.  Key pieces of 
information, such as the metallicity and ionization of the 
HVCs, have remained elusive. This situation has improved in recent years 
as absorption line observations toward quasars and  active galactic nuclei 
have been conducted with the {\it Hubble Space Telescope} (HST).
For example, Lu \et (1998) have found that HVC\,287+22+240 has a metallicity 
(S/H)\,$\sim$\,0.25(S/H)$_\odot$, with a dust to gas ratio traced by 
(S/Fe) similar to that in the Magellanic Clouds.  This determination 
rests critically upon the assumption that the ionization correction
for the amounts of S\,{\sc ii} and Fe\,{\sc ii} arising in ionized gas 
associated with the HVCs are small [i.e., 
N(S)/N(Fe)/N(H)\,$\approx$\,N(S\,{\sc ii})/N(Fe\,{\sc ii})/N(\hi)].
Using HST data, Wakker and collaborators (this volume) 
have found (S/H)\,$\sim$\,0.1(S/H)$_\odot$ for Complex~C in the 
direction of Mrk\,290.  This determination appears fairly robust, as 
additional information about ionized gas is available from H$\alpha$ imaging 
of the sight line.  The most complete set of HST measurements for studying
ionization in high velocity clouds exists for the ``\chvcs'' toward Mrk\,509 
(Sembach \et 1999, this volume). 

\begin{figure}[!ht]
\includegraphics{sembach1_fig2.ps}
\vspace{6.6in}
\footnotesize \noindent 
{\small Figure~2.} - Curve of growth for \oi\ (top), 
\ni\ (middle), and \feii\ (bottom) lines in the FUSE bandpass.  The data
points are appropriate for a single component model with a Doppler width
of 5 \kms, N(\hi) = 1$\times$10$^{20}$ atoms cm$^{-2}$, and solar
abundances.   Weak lines that are badly blended 
with other lines are not shown. The dashed lines are curves for
Doppler widths of 3 \kms\ (lower curve) and 10 \kms\ (upper curve).  
The inset horizontal tick marks in
the \feii\ panel indicate the values of log\,(W$_\lambda$/$\lambda$)
for $\tau_0$ = 0.3, 1.0, and 3.0. 
\end{figure}

\subsection{Abundances}
FUSE observations will provide valuable information for
determining the metallicities and ionization of HVCs. In particular,
measurements of \oi, which has an ionization potential of 13.6 eV
and is strongly tied to \hi\ through charge exchange reactions
(Spitzer 1978), will yield oxygen abundances for neutral clouds.  
Reliable oxygen abundances currently exist for very few clouds, since
the two primary transitions in the HST wavelength region are either
heavily saturated in most directions ($\lambda$1302.168) or extremely
weak ($\lambda$1355.598) (see Meyer, Jura, \& Cardelli 1998).  

Besides containing \oi\ lines, the FUSE bandpass encompasses 
lines of abundant elements that 
will be useful for estimating gas-phase abundances in the neutral 
ISM and HVCs.  Several atomic species have lines spanning a large range in 
$f\lambda$ (e.g., \oi, \ni, and \feii).  Curves of growth for these species
are shown in Figure~2 for N(\hi)\,=\,1$\times$10$^{20}$ atoms cm$^{-2}$ and
no gas-phase depletion onto dust.  For other values of N(\hi) and depletions,
D, the lines move horizontally along these curves by an amount equal to 
log\,N(\hi)\,+\,D\,--\,20.
Additional low ionization lines of \cii, \mgii, \alii, Si\,{\sc ii}, \pii, and
\ari\ are plotted on the curve of
growth shown in Figure~3.  The strong line of  \cii\ $\lambda$1036 and
the stronger \oi\ and \ni\ lines shown in Figure~2
are likely to be heavily saturated in their cores along
many sight lines but can be used to trace low density, high velocity 
dispersion gas in their absorption wings.  Weak \mgii\ lines near 
1026\AA\ will be difficult to recover in the wings of \hi\ Ly$\beta$
and are not shown in Figure~3. For information about local ISM 
abundances derived from {\it Copernicus} data, see 
Jenkins, Savage, \& Spitzer (1986) and Jenkins (1987, and references therein).

\begin{figure}[!h]
\includegraphics{sembach1_fig3.ps}
\vspace{2.4in}
\footnotesize \noindent 
{\small Figure~3.} Curves of growth for far-UV lines of various heavy
element species not shown in Figure~2.  Values for \cli\ and \clii\
are not shown.  The data
points are appropriate for a single component model with a Doppler width
of 5 \kms, N(\hi) = 1$\times$10$^{20}$ atoms cm$^{-2}$, and solar
abundances.  The dashed lines are curves for
Doppler widths of 3 \kms\ (lower curve) and 10 \kms\ (upper curve).  
\end{figure}

The primary obstacle
facing abundance studies of HVCs is the difficulty in making accurate 
determinations of neutral hydrogen column densities through 21\,cm emission 
since some HVCs appear to contain structure at arc minute scales (c.f., 
Wakker \& van~Woerden 1997).  FUSE observations of higher order Lyman series 
lines of \hi\ will probably not address this problem, except in special 
circumstances.  It might be possible to do so when the velocity of the HVC
is large ($|$\vlsr$|$\,$\sim$\,200 \kms), the width of the low velocity
absorption is small (FWHM\,$<$\,50 \kms), and the amount of intermediate 
velocity gas is negligible (N(\hi)\,$<$\,few$\times$10$^{16}$ atoms cm$^{-2}$).
Detections in multiple \hi\ lines will be
necessary since a typical HVC \hi\ column density 
of $\sim$10$^{19}$ atoms cm$^{-2}$
will place most of the \hi\ lines on the flat part of the curve of growth.

\subsection{Ionization}

Absorption line observations of ionized gas species in some
HVCs indicate that they can contain large quantities of ionized gas 
(Sembach \et 1995, 1999, this volume).  H$\alpha$ imaging of the larger
HVC complexes (Tufte, Reynolds, \& Haffner 1998, this volume) also reveals the 
presence of ionized gas.  Therefore, it is important to consider 
whether gas-phase abundances derived from measurements of singly charged 
ions (e.g., \siii, \pii, \feii) or neutral atoms with first ionization
potentials greater than 13.6 eV (e.g., \ni, \ari) are affected by
\hii\ region contributions. 
Lines of adjacent ionization stages of many
elements in the far-UV (e.g., C\,{\sc i}-{\sc iii}, N\,{\sc i}-{\sc iii}, 
P\,{\sc ii}-{\sc v}, S\,{\sc iii}-{\sc iv}, Cl\,{\sc i}-{\sc ii},
Ar\,{\sc i}-{\sc ii}, and Fe\,{\sc ii}-{\sc iii}) can be used to 
estimate ionization corrections when necessary. 

Ionized gas diagnostics in the far-UV wavelength region span a large range in
ionization potential.  Of these,
the strong \ovi\ lines are the most important for studying gases with 
T\,$\ge$\,10$^5$ K.  The observable lines of other elements trace 
lower temperature gases ranging from 10$^4$ to $\sim$10$^5$ K.  Table~1
contains a list of the ionized gas resonance lines in the FUSE bandpass.
Lines of species having creation ionization potentials greater
than 13.6 eV are included.  For each ion, predicted line strengths
and widths are listed for a simple model in which a solar metallicity 
gas with no dust and  N(\hii)\,=\,1x10$^{20}$~cm$^{-2}$ is in 
collisional ionization equilibrium.  In non-equilibrium situations, the 
ionization fractions will differ from those listed (see 
Sutherland \& Dopita 1993).  

Some of the lines listed in Table~1 are very strong and will be heavily
saturated even when N(\hii) is low.  For example, 
$\tau_0$(\ciii\ $\lambda$977)\,$>$\,3 when
N(\hii) $>$ few\,$\times$\,10$^{16}$ atoms cm$^{-2}$.  Lines of other
ions, such as \piii, \pv, and \criii, will be
weak even when N(\hii) is large.  Incorporation of the 
refractory elements (e.g., Cr, Fe) into dust will affect the observed line 
strengths of \criii\ and \feiii.

\begin{table}[!h]
\caption{\bf Ionized Gas Resonance Lines in the FUSE Bandpass\tablenotemark{a}}
\begin{center}
\footnotesize
\begin{tabular}{lrrcrcrccc}
\tableline
\tableline
Ion & \multicolumn{1}{c}{Z} & \multicolumn{1}{c}{A} 
& \multicolumn{1}{c}{IP$^{(i, i+1)}$}
& \multicolumn{1}{c}{T$_{max}$} & \multicolumn{1}{c}{f$_{ion}$}
& \multicolumn{1}{c}{$\lambda$} & \multicolumn{1}{c}{log~$f\lambda$}
& \multicolumn{1}{c}{$\tau_0$}  & \multicolumn{1}{c}{b$_0$}\\
&  &   & \multicolumn{1}{c}{(eV)}
& \multicolumn{1}{c}{(K)} & & \multicolumn{1}{c}{(\AA)}
&  &  & \multicolumn{1}{c}{(\kms)}\\
\tableline
\ciii  &  6 & 8.55 & 24.38, 47.89 & 70,000 & 0.832 &  977.020 & 2.872 & 3340 & \phn9.8\smallskip
\\
\nii   &  7 & 7.97 & 14.53, 29.60 & 25,000 & 0.971 &  915.612 & 2.123 & 330  & \phn5.4\\
       &    &       &              &        &       & 1083.990 & 2.048 & 278  & \phn5.4\smallskip
\\
\niii  &  7 & 7.97 & 29.60, 47.45 & 80,000 & 0.769 &  989.799 & 2.023 & 116  & \phn9.7\smallskip
\\
\ovi   &  8 & 8.87 & 113.9, 138.1 &280,000 & 0.220 & 1031.926 & 2.137 & 196  & 17.1\\
       &    &       &              &        &       & 1037.617 & 1.836 &  98  & 17.1\smallskip
\\
\piii  & 15 & 5.57 & 19.72, 30.18 & 35,000 & 0.792 &  998.000 & 2.047 &  1.1 & \phn4.3\smallskip
\\
\piv   & 15 & 5.57 & 30.18, 51.37 & 70,000 & 0.706 &  950.657 & 3.044 &  7.1 & \phn6.1\smallskip
\\
\pv    & 15 & 5.57 & 51.37, 65.02 &100,000 & 0.609 & 1117.977 & 2.723 &  2.4 & \phn7.3\\
       &    &       &              &        &       & 1128.008 & 2.422 &  1.2 & \phn7.3\smallskip
\\
\siii  & 16 & 7.27 & 23.33, 34.83 & 50,000 & 0.838 & 1012.502 & 1.556 & 16.5 & \phn5.1\smallskip
\\
\siv   & 16 & 7.27 & 34.83, 47.30 &100,000 & 0.610 & 1062.662 & 1.628 & 10.0 & \phn7.2\smallskip
\\
\svi   & 16 & 7.27 & 72.68, 88.05 &180,000 & 0.140 &  933.378 & 2.319 & 16.6 & \phn9.7\\
       &    &       &              &        &       &  944.523 & 2.615 &  8.3 & \phn9.7\smallskip
\\ 
\arii  & 18 & 6.56 & 15.76, 27.63 & 22,000 & 0.964 &  919.781 & 0.912 &  1.4 & \phn3.0\smallskip
\\
\criii & 24 & 5.68 & 16.50,30.96  & 28,000 & 0.893 &  923.780 & 1.874 &  1.6 & \phn3.0\\
       &    &       &              &        &       & 1030.100 & 1.809 &  1.4 & \phn3.0\\
       &    &       &              &        &       & 1033.331 & 1.820 &  1.4 & \phn3.0\\
       &    &       &              &        &       & 1040.050 & 2.104 &  2.7 & \phn3.0\smallskip
\\ 
\feiii & 26 & 7.51 & 16.18, 30.65 & 28,000 & 0.893 & 1122.526 & 1.947 & 132 & \phn2.9\\
\tableline
\end{tabular}
\end{center}
\scriptsize
\tablenotetext{a}{This table contains information for ionized gas lines
in the 905--1187\AA\ wavelength region.  Fine structure transitions of \nii\
and \niii\ occur at wavelengths within 1--2\AA\ of the \nii\ and \niii\ 
lines listed in this table.
Columns 2--9 are as follows:\\
\begin{tabular}{ll}
Z: & Atomic number. \\
A: & Solar abundance of element (total of all ions) relative to H on a 
logarithmic\\
 & scale where A(H) = 12.00.\\
IP: & Ionization potential to create and destroy the listed ion 
(Moore 1971). \\
T$_{max}$: & Temperature at which ion peaks in abundance in collisional 
ionization equilibrium. \\
& Values are from Sutherland \& Dopita (1993), except for P and Cr.  
The values for P \\
& were obtained by interpolating along iso-electronic sequences of adjoing 
even-Z \\
& elements (Si, S). For \criii, a value equal to those for \feiii\ and 
Ni\,{\sc iii} was adopted. \\
f$_{ion}$: & Fractional abundance of ion in collisional ionization at 
temperature T$_{max}$. \\
$\lambda$: & Wavelength from Morton (1991). \\
log~$f\lambda$: & Product of wavelength (in \AA) and f-value from Morton 
(1991). \\
$\tau_0$: & Optical depth of line at line center for a plasma with 
N(\hii) = 1x10$^{20}$~cm$^{-2}$,\\
& assuming collisional ionization equilibrium, the listed values of A, 
T$_{max}$, f$_{ion}$, \\
& and $b_0$, solar metallicity, and no gas-phase depletion due to dust.\\
$b_0$: & Thermal broadening parameter of line at temperature T$_{max}$.
\end{tabular}}
\end{table}

\subsection{Molecules}

FUSE will be able to search for dust and molecules in HVCs. Comparisons
of Fe to O and other lightly depleted elements will reveal whether the gas
has a solar abundance pattern, independent of whether the \hi\ 
column density is known.  Searches for H$_2$ absorption will reveal
columns as small as 10$^{14}$ molecules cm$^{-2}$, or about 5 orders 
of magnitude lower than has been possible with millimeter 
wavelength observations of CO (e.g., Wakker \et 1997).

\subsection{FUSE Team Observations of HVCs}
Observations of high velocity clouds (HVCs) play prominent 
roles in the FUSE D/H and \ovi\ programs.  Table~2 contains a list of 
selected HVCs toward objects that will be observed in detail by the FUSE 
Science Team.  The Galactic coordinates of the background sources, 
HVC identifications, and velocities of the HVCs are listed.  
The extragalactic sight lines listed currently have 
planned observations sufficient to produce S/N\,$\ge$\,10 per FUSE 
resolution element.   An additional $\sim$50--100 extragalactic 
sight lines will be inspected at low resolution to determine far-UV flux 
levels and sight line velocity structure.  Many of these ``snapshot'' 
sight lines pass through or near Complexes A, C, and M.  The three
Galactic sight lines listed have HVCs that have been observed in 
absorption and represent a small subset of the total number of Galactic
sight lines that will be observed.
 
\begin{table}[!h]
\caption{\bf Selected High Velocity Cloud Sight Lines in the FUSE 
Science Team Program\tablenotemark{a}}
\begin{center}
\small
\begin{tabular}{llrrc}
\tableline
\tableline
Object & HVC Name & $l$(\degr) & $b$(\degr) & \vlsr(\kms) \\
\tableline
\multicolumn{5}{c}{Extragalactic Sight Lines\tablenotemark{a}}\\
\tableline
PKS\,2155-304  	& \chvcs	& 17.7 & $-$52.2 & $-$140, $-$256 \\
Mrk\,509	& \chvcs	& 36.0 & $-$29.9 & $-$228, $-$283 \\
Mrk\,290	& Complex~C	& 91.5 & +48.0 & $-$136 \\
H\,1821+643	& Outer Arm	& 94.0 & +27.4 & $-$120 \\
Mrk\,817	& Complex~C	& 100.3 & +53.5 & $-$107 \\
NGC\,3783	& HVC~287+22+240 & 287.5 & +23.0 & +240 \\
Fairall\,9 	& Magellanic Stream & 295.1 & $-$57.8 & +170, +210 \\
\tableline
\multicolumn{5}{c}{Galactic Sight Lines\tablenotemark{b}} \\
\tableline
LS\,4825	& Inner Galaxy  & 1.7  & $-$6.6 & $-$206, $-$150, +93\\
BD\,+38~2182    & Complex~M	& 182.2 & +62.2 & $-$93 \\
HD\,156359      & Uncatalogued  & 328.7 & $-$14.5 & +125 \\
\tableline
\end{tabular}
\end{center}
\footnotesize
\tablenotetext{a}{Observations of stars in the Large and Small Magellanic 
Clouds will also provide information on high velocity gas in those directions.
Additional HVCs along extragalactic sight lines may be 
explored through the D/H snapshot program, which will produce short 
exposures of 50--100 extragalactic objects.}
\tablenotetext{b}{These are but a few of the many Galactic sight lines 
that will be observed by FUSE.  This table does not include the numerous 
intermediate velocity clouds or high velocity gas features associated with 
known supernova remnants (e.g., Vela) or star-forming regions (e.g., Carina) 
that will be observed. Approximately 200 sight lines in the 
Galactic disk and halo will be observed as part of the D/H and \ovi\ programs}
\end{table}

\section{Simulated Spectra for Simple \hi\ Cloud Models}

Given the large number of atomic and molecular transitions in the far-UV
suitable for studies of HVCs, it is instructive to consider the
absorption signatures expected for simple interstellar cloud properties.
These results can then be applied to more complicated situations. 
 
\subsection{Absorption Lines Viewed at FUSE Resolution} 

The {\it apparent} optical depth of a spectral line at a velocity $v$ is
given by $\tau_a(v)$ = ln($I_c(v)/I_o(v)$), where $I_o$ and $I_c$ are the
observed and continuum (unattenuated) intensities, respectively.
This differs from the true optical depth, $\tau(v)$ = ln($I_c(v)/I(v)$),
due to the finite resolution of the spectral spread function 
of the instrument used to observe the absorption line.  The effect of 
this convolution on the line shape of a single Gaussian component 
is shown in Figure~4 for three values of the Doppler width $b_0$ and 
central optical depth $\tau_0$ of the line.
An 
instrumental spread function width appropriate for FUSE,
$b_I$\,=\,(c$\lambda/\Delta\lambda$)/(2$\sqrt{ln 2}$)\,$\approx$\,
6 \kms, has been applied.
The more severely a line is under-resolved by the instrument, 
the greater the difference between the true optical depth of the line and its 
apparent optical depth.  The resulting 
unresolved saturated structure must be accounted for
in determinations of the column density contained within the line.  
Detailed discussions of apparent optical depths and the derivation of 
column densities from apparent optical depth profiles have been given by 
Savage \& Sembach (1991) and Jenkins (1996).

\begin{figure}[!h]
\includegraphics{sembach1_fig4.ps}
\vspace{1.5in}
\footnotesize \noindent 
{\small Figure~4.} Line profiles of intrinsic widths $b_0$ and 
central optical depths  $\tau_0$ ({\it left}:~$\tau_0$\,=\,0.3, 
{\it middle}: $\tau_0$\,=\,1.0, {\it right}: $\tau_0$\,=\,3.0)
convolved with a Gaussian line spread function having an instrumental
resolution appropriate 
for FUSE, ($\lambda/\Delta\lambda$)\,$\approx$\,30,000.
\end{figure}

\subsection{Model~1: A Sight Line Through the Neutral Medium of the 
Galactic Halo}

Figure~5 contains a noiseless simulation of the absorption expected in the
910--1160 \AA\ spectral region for  
a single interstellar cloud containing 3$\times$10$^{20}$ atoms cm$^{-2}$ of
\hi\ and 1$\times$10$^{17}$ molecules~cm$^{-2}$ of H$_2$.  The cloud has a 
temperature of 100 K, a b-value (thermal + turbulent) of 5 \kms, and 
a warm halo cloud gas-phase depletion pattern 
(Savage \& Sembach 1996).\footnotemark\ \footnotetext{
Gas-phase depletions for Mg, Al, Si, S, Cr, Mn, Fe, and Ni are included.
[Mg/H]\,=\,$-$0.50, [Al/H]\,$\equiv$\,[Fe/H]\,=\,$-$0.60, 
[Si/H]\,=\,$-$0.50, [S/H]\,=\,0.0, [Cr/H]\,=\,$-$0.50, [Mn/H]\,=\,$-$0.60, 
[Ni/H]\,=\,$-$0.85.  All other elements (X) are assumed to have solar
abundances, [X/H]\,=\,log(X/H)$-$log(X/H)$_\odot$\,=\,0.0. } 
The model includes lines having 
$\tau_0$\,$>$\,0.05 and an instrumental resolution appropriate for FUSE.
Lines of atomic species and H$_2$ are indicated at the top of each panel.
The D/H ratio used is the local ISM value of 1.6$\times$10$^{-5}$ 
(Linsky \et 1993).
For elements with first ionization potentials below 13.6 eV, 
it is assumed
that 0.5\% of the elemental gas-phase abundance is neutral, with
the remaining 99.5\% being singly ionized.\footnotemark\ Chlorine is an 
exception since reactions with H$_2$ can convert \clii\ to \cli\
(Jura 1974; Jura \& York 1978); it is assumed that 20\% of the Cl is \cli\ and 
80\% is \clii, which is typical of low density regions with large values
of N(\hi)/N(H$_2$) (see Harris \& Bromage 1984).  Table~3 contains a summary
of the model parameters.

\footnotetext{This is an approximate value. The relative amounts of 
neutral and singly ionized species will depend upon the detailed physical 
conditions of the cloud.}

The richness of the far-UV wavelength region shown in Figure~5 and the 
progression to higher line densities as wavelength
decreases to the Lyman limit at 912\AA\ is striking.  Even for the simple,
single component model shown, line blending can be problematic, especially
at shorter wavelengths.  

\begin{table}[!h]
\caption{\bf Model Parameters for Simulated Spectra\tablenotemark{a}}
\begin{center}
\footnotesize
\begin{tabular}{lcc}
\tableline
\tableline
Parameter & Model~1 & Model~2 \\
& (Figure~5) & (Figure~6) \\
\tableline
N(\hi) (cm$^{-2}$)	& 3$\times$10$^{20}$ 	& 1$\times$10$^{20}$  \\
N(H$_2$) (cm$^{-2}$)	& 1$\times$10$^{17}$ 	& 1$\times$10$^{17}$  \\
$b_0$ (\kms)		& 5 	 	& 5 \\
$v_0$ (\kms)		& 0		& 0 \\
T (K)			& 100 		& 500  \\	
Metallicity 		& Solar  	& Solar \\
Dust			& Warm halo	& None \\
D/H ratio		& 1.6$\times$10$^{-5}$	& 1.6$\times$10$^{-5}$ \\
$b_I$ (\kms) 		& 5.9 & 5.9 \\
$\tau_{thresh}$		& 0.05		& 0.05 \\
No. of lines (atomic, H$_2$)	& 302, 190	& 278, 393\\
\tableline
\end{tabular}
\end{center}
\footnotesize
\tablenotetext{a}{The models shown in Figures~5 and 6 are appropriate for a 
single interstellar cloud with the parameters listed in this table.  
The solar abundances used are meteoritic
values from Anders \& Grevesse (1989), except for C, N, and O, which are
photospheric values from Grevesse \& Noels (1993).  The Savage \& Sembach
(1996) warm halo cloud gas-phase depletion pattern due to dust 
is used in Model 1; elements not included in their study are assumed to have 
solar abundances, except for Al, which is set to the value for Fe and Ni.
The atomic and H$_2$ data in these models are from Morton (1991) and
Abgrall \et (1993a, 1993b).}
\end{table}

Atomic fine-structure lines and molecular lines of HD and CO are not included 
in this model.  Information about additional atomic lines and CO can be found 
in the data compilations presented by Morton (1991) and Morton \& Noreau 
(1994). Table~4 contains a brief list of some of the transitions
in the Lyman ($B^1\Sigma^+_u-X^1\Sigma^+_g$) and Werner 
($C^1\Pi_u-X^1\Sigma^+_g$) systems of HD.  The wavelengths are from 
Dabrowski \& Herzberg (1976).  The f-values for these lines were computed
using the band oscillator strengths calculated by Allison \& Dalgarno (1970).
Some of these transitions were seen in the spectrum of 
$\zeta$~Ophiuchi by {\it Copernicus} (Wright \& Morton 1979).

\subsection{Model~2: An \hi\ Cloud with Warm H$_2$}

Figure~6 contains a second sample spectrum of the same cloud in Figure~5 with
a lower value of N(\hi), a temperature of 500 K, and no gas-phase depletion
onto dust grains. Other model parameters are summarized in Table~3. 
In this case, higher order (J\,$\ge$\,3) rotational lines of H$_2$ blanket 
the spectrum.

The models shown in Figures~5 and 6 are intended to serve as aids in 
identifying lines of various elemental species in the FUSE wavelength
range.  These figures can be used to estimate line strengths for different 
cloud parameters through the following scaling relation:
\begin{center}
$\tau_0$ $\propto$ $\frac{N(\hi)}{b_0}$ 10$^{(A+D)}$,
\end{center}
where $\tau_0$ is the optical depth at line center, N(\hi) is the column 
density
of \hi, $b_0$ is the Doppler spread parameter for the line, A is
the logarithmic abundance of the element on a scale where 
A(H)\,=\,12.00, and D is the logarithmic gas-phase depletion of the element
relative to the reference abundances used.  Note 
that $\tau_0$ is the true optical depth of the line, not the {\it apparent}
optical depth, $\tau_a$, which is less than $\tau_0$ in situations where the 
instrumental line function width, $b_I$, is greater than the intrinsic width 
of the line, $b_0$. For most neutral gas species observed by FUSE,  
$b_I$\,$>$\,$b_0$, and therefore $\tau_a$~$<$~$\tau_0$.  Figure~4 can
be used to relate $\tau_a$ and $\tau_0$ for FUSE data.

\begin{table}[!ht]
\caption{\bf Selected HD Lines in the FUSE Bandpass\tablenotemark{a}}
\begin{center}
\footnotesize
\begin{tabular}{llccllc}
\tableline
\tableline
   Transition & \multicolumn{1}{c}{$\lambda$(\AA)} & f-value 
&& Transition & \multicolumn{1}{c}{$\lambda$(\AA)} & f-value\\
\tableline
\multicolumn{7}{c}{$B^1\Sigma^+_u-X^1\Sigma^+_g$ (Lyman series)} \\
\tableline
0--0 R(0)  & 1105.838$^*$  & 7.60 ($-$4) && 1--0 R(0)  & 1091.999$^*$  & 2.99 ($-$3) \\
0--0 R(1)  & 1106.214$^*$  & 5.07 ($-$4) && 1--0 R(1)  & 1092.397$^*$  & 1.98 ($-$3) \\
0--0 P(1)  & 1107.289$^*$  & 2.53 ($-$4) && 1--0 P(1)  & 1093.400$^*$  & 9.95 ($-$4) \\
0--0 R(2)  & 1107.325$^*$  & 4.56 ($-$4) && 1--0 R(2)  & 1093.526$^*$  & 1.79 ($-$3) \\
0--0 P(2)  & 1109.114$^*$  & 3.03 ($-$4) && 1--0 P(2)  & 1095.194$^*$  & 1.19 ($-$3) \smallskip
\\
2--0 R(0)  & 1078.828$^*$  & 6.74 ($-$3) && 3--0 R(0)  & 1066.271  & 1.14 ($-$2) \\
2--0 R(1)  & 1079.242$^*$  & 4.49 ($-$3) && 3--0 R(1)  & 1066.708  & 7.59 ($-$3) \\
2--0 P(1)  & 1080.182$^*$  & 2.24 ($-$3) && 3--0 P(1)  & 1067.585$^*$  & 3.79 ($-$3) \\
2--0 R(2)  & 1080.381$^*$  & 4.04 ($-$3) && 3--0 R(2)  & 1067.842  & 6.82 ($-$3) \\
2--0 P(2)  & 1081.946$^*$  & 2.69 ($-$3) && 3--0 P(2)  & 1069.319  & 4.54 ($-$3) \smallskip
\\
4--0 R(0)  & 1054.288  & 1.61 ($-$2) && 5--0 R(0)  & 1042.847  & 2.01 ($-$2) \\
4--0 R(1)  & 1054.722$\dagger$  & 1.07 ($-$2) && 5--0 R(1)  & 1043.288  & 1.34 ($-$2) \\
4--0 P(1)  & 1055.563  & 5.36 ($-$3) && 5--0 P(1)  & 1044.082  & 6.68 ($-$3) \\
4--0 R(2)  & 1055.877  & 9.64 ($-$3) && 5--0 R(2)  & 1044.442  & 1.20 ($-$2) \\
4--0 P(2)  & 1057.266  & 6.42 ($-$3) && 5--0 P(2)  & 1045.759  & 8.00 ($-$3) \smallskip
\\
6--0 R(0)  & 1031.912  & 2.28 ($-$2) && 7--0 R(0)  & 1021.456  & 2.42 ($-$2) \\
6--0 R(1)  & 1032.361  & 1.52 ($-$2) && 7--0 R(1)  & 1021.916  & 1.62 ($-$2) \\
6--0 P(1)  & 1033.114  & 7.60 ($-$3) && 7--0 P(1)  & 1022.626  & 8.07 ($-$3) \\
6--0 R(2)  & 1033.514  & 1.37 ($-$2) && 7--0 R(2)  & 1023.064  & 1.45 ($-$2) \\
6--0 P(2)  & 1034.764  & 9.11 ($-$3) && 7--0 P(2)  & 1024.249  & 9.67 ($-$3) \smallskip
\\
8--0 R(0)  & 1011.457$\dagger$  & 2.44 ($-$2) && 9--0 R(0)  & 1001.892  & 2.35 ($-$2) \\
8--0 R(1)  & 1011.924  & 1.63 ($-$2) && 9--0 R(1)  & 1002.360  & 1.57 ($-$2) \\
8--0 P(1)  & 1012.590$\dagger$  & 8.12 ($-$3) && 9--0 P(1)  & 1003.003  & 7.84 ($-$3) \\
8--0 R(2)  & 1013.074  & 1.46 ($-$2) && 9--0 R(2)  & 1003.507  & 1.41 ($-$2) \\
8--0 P(2)  & 1014.195  & 9.73 ($-$3) && 9--0 P(2)  & 1004.580  & 9.39 ($-$3) \\
\tableline
\multicolumn{7}{c}{$C^1\Pi_u-X^1\Sigma^+_g$ (Werner series)} \\
\tableline
0--0 R(1)  & 1007.251$\dagger$  & 1.73 ($-$2) && 1--0 R(1)  &  \phn987.276$\dagger$  & 3.07 ($-$2) \\
0--0 R(0)  & 1007.283$\dagger$  & 3.45 ($-$2) && 1--0 R(0)  &  \phn987.276$\dagger$  & 6.14 ($-$2) \\
0--0 R(2)  & 1007.650  & 1.38 ($-$2) && 1--0 R(2)  &  \phn987.712  & 2.45 ($-$2) \\	
0--0 Q(1)  & 1008.199$\dagger$  & 1.73 ($-$2) && 1--0 Q(1)  &  \phn988.145  & 3.07 ($-$2) \\ 
0--0 Q(2)  & 1009.080  & 1.73 ($-$2) && 1--0 Q(2)  &  \phn989.021  & 3.06 ($-$2) \\
0--0 P(2)  & 1010.005$\dagger$  & 3.45 ($-$3) && 1--0 P(2)  &  \phn989.893  & 6.12 ($-$3) \smallskip
\\
2--0 R(0)  & \phn968.972  & 6.63 ($-$2) && 3--0 R(0)  &  \phn952.208  & 5.72 ($-$2) \\
2--0 R(1)  & \phn969.030  & 3.31 ($-$2) && 3--0 R(1)  &  \phn952.285  & 2.86 ($-$2) \\
2--0 R(2)  & \phn969.550  & 2.65 ($-$2) && 3--0 R(2)  &  \phn952.802  & 2.29 ($-$2) \\	
2--0 Q(1)  & \phn969.822  & 3.31 ($-$2) && 3--0 Q(1)  &  \phn953.046  & 2.86 ($-$2) \\ 
2--0 Q(2)  & \phn970.697  & 3.31 ($-$2) && 3--0 Q(2)  &  \phn953.945$\dagger$  & 2.85 ($-$2) \\
2--0 P(2)  & \phn971.490  & 6.61 ($-$3) && 3--0 P(2)  &  \phn954.637$\dagger$  & 5.70 ($-$3) \\
\tableline
\end{tabular}
\end{center}
\tablenotetext{a}{Vaccum wavelengths in this table are from the wavenumbers 
measured by Dabrowski \& Herzberg (1976).  A dagger ($\dagger$) next to the  
wavelength indicates that the line may be blended in their spectrum.  An
asterisk (*) indicates that the wavelengths are calculated from the rotational
constants and band origins listed by Dabrowski \& Herzberg (1976).
The f-values are based upon the
band oscillator strengths calculated by Allison \& Dalgarno (1970).  
A value of 1.23 (-4) is equivalent to 1.23$\times$10$^{-4}$.
}
\end{table}

\begin{figure}[!ht]
\includegraphics{sembach1_fig5a.ps}
\vspace{7.3in}
\scriptsize \noindent 
{\small Figure 5a.} - Model spectrum \#1 (\hi\ halo cloud sight line).  
The parameters for this model are 
listed in Table~3.  Vertical dashed lines indicate the wavelengths 
of the ionized gas lines listed in Table~1. Fine structure lines of 
\ci, \cii, \nii, and \niii\ are not shown.  The inset box shows a Poisson 
noise level S/N\,=\,17 {\it per FUSE pixel}, which corresponds to 
S/N\,=\,30 {\it per resolution element} (3 pixels). The
horizontal error bar in the lower left of each panel indicates a 
velocity range of 100 \kms.
\end{figure}

\clearpage
\begin{figure}[!t]
\includegraphics{sembach1_fig5b.ps}
\vspace{7.3in}
\scriptsize \noindent 
{\small Figure 5b.} Same as Figure 5a, except for the 960--1010 \AA\
wavelength region.
\end{figure}

\clearpage
\begin{figure}[!t]
\includegraphics{sembach1_fig5c.ps}
\vspace{7.3in}
\scriptsize \noindent 
{\small Figure 5c.} Same as Figure 5a, except for the 1010--1060 \AA\
wavelength region.
\end{figure}

\clearpage
\begin{figure}[!t]
\includegraphics{sembach1_fig5d.ps}
\vspace{7.3in}
\scriptsize \noindent 
{\small Figure 5d.} Same as Figure 5a, except for the 1060--1110 \AA\
wavelength region.
\end{figure}

\clearpage
\begin{figure}[!t]
\includegraphics{sembach1_fig5e.ps}
\vspace{7.3in}
\scriptsize \noindent 
{\small Figure 5e.} Same as Figure 5a, except for the 1110--1160 \AA\
wavelength region.
\end{figure}

\clearpage
\newpage
\begin{figure}[!t]
\includegraphics{sembach1_fig6a.ps}
\vspace{7.3in}
\scriptsize \noindent 
{\small Figure 6a.} Model spectrum \#2 (\hi, warm H$_2$ sight line).  
The parameters for this model are 
listed in Table~3.  Vertical dashed lines indicate the wavelengths 
of the ionized gas lines listed in Table~1. Fine structure lines of 
\ci, \cii, \nii, and \niii\ are not shown. The
horizontal error bar in the lower left of each panel indicates a 
velocity range of 100 \kms.
\end{figure}

\clearpage
\begin{figure}[!t]
\includegraphics{sembach1_fig6b.ps}
\vspace{7.3in}
\scriptsize \noindent 
{\small Figure 6b.} Same as Figure 6a, except for the 960--1010 \AA\
wavelength region.
\end{figure}

\clearpage
\begin{figure}[!t]
\includegraphics{sembach1_fig6c.ps}
\vspace{7.3in}
\scriptsize \noindent 
{\small Figure 6c.} Same as Figure 6a, except for the 1010--1060 \AA\
wavelength region.
\end{figure}

\clearpage
\begin{figure}[!t]
\includegraphics{sembach1_fig6d.ps}
\vspace{7.3in}
\scriptsize \noindent 
{\small Figure 6d.} Same as Figure 6a, except for the 1060--1110 \AA\
wavelength region.
\end{figure}

\clearpage
\begin{figure}[!t]
\includegraphics{sembach1_fig6e.ps}
\vspace{7.3in}
\scriptsize \noindent 
{\small Figure 6e.} Same as Figure 6a, except for the 1110--1160 \AA\
wavelength region.
\end{figure}

\clearpage
\newpage

\section{Science Team Members}
FUSE is a Principal Investigator class mission with a Science Team 
composed 
of U.S., Canadian, and French scientists at academic and 
government institutions.  
The Principal Investigator of the FUSE mission is H. Warren Moos of 
the Johns Hopkins University.  
Members of the FUSE Science Team include: Webster Cash, 
Lennox Cowie, Arthur Davidsen, 
Andrea Dupree, Paul Feldman, Scott Friedman, James Green, Richard Green,
Cecile Gry (associate), John Hutchings, Edward Jenkins, Jeffrey Linsky, 
Roger Malina, Blair Savage,
J. Michael Shull, Oswald Siegmund, George Sonneborn, Theodore Snow, 
Alfred Vidal-Madjar, Alan Willis (associate), Bruce Woodgate, and Donald York.

In addition to the Science Team, there are numerous members of the 
FUSE Instrument and Operations Teams at the Johns Hopkins University, the 
University of Colorado, and the University of California who have contributed
to the instrumental development, mission planning, science planning, 
and science operations of FUSE.  These scientists will actively participate
in the analysis of data obtained with Science Team observing time.

\section{The FUSE Satellite}
The FUSE satellite is composed of a three-axis stabilized spacecraft and 
the scientific instrument.  The total satellite weight is 1360 kg.  
The instrument consists of four co-aligned 
telescopes optimized for transmission at far-UV wavelengths. The light 
from the four channels is dispersed by four spherical, aberration-corrected 
holographic diffraction gratings.  Two channels with SiC coatings 
cover 905--1100\AA, and two channels with Al+LiF coatings cover 
1000--1187\AA.  Two
delay-line microchannel plate detectors each detect one SiC and one 
Al+LiF channel.

Wavelength overlap, physical separation of
the four channels, and complete wavelength coverage on each detector
provides for high sensitivity and redundancy.  The instrument also 
contains a fine error sensor to identify the pointing location and to 
stabilize the spacecraft during observations.
Properties of the instrument are summarized in Table~5.
 
\begin{table}[h!]
\caption{Instrument Parameters}
\begin{center}
\small
\begin{tabular}{ll}
\tableline
\tableline
Mirrors (4):& Off-axis parabolas, zerodur substrate, 387$\times$352 mm
clear \\
& aperture\\
Gratings (4): & Spherical, aberration-corrected, holographically ruled,\\
& characteristic line densities of 5767 $l$/mm (SiC) and \\
& 5350 $l$/mm (LiF)\\
Optics Coatings: & SiC or Al+LiF \\
Detectors (2): & Microchannel plates with double delay-line anodes \\
& and KBr photocathodes \\
Spectrograph Design: & 1.652 m Rowland circle \\
Instrument Size: & 1.2 m $\times$ 1.8 m $\times$ 4.4 m \\
Instrument Mass: & 780 kg \\
\tableline
\end{tabular}
\end{center}
\end{table}

Table~6 contains pre-launch predictions for the in-orbit performance 
of FUSE.  These quantities will be updated as in-orbit activities 
progress and astronomical observations are obtained.
Figure~7 contains a plot of effective area at the beginning of the mission
versus wavelength.
 
\begin{table}[!h]
\caption{FUSE Predicted Performance}
\begin{center}
\small
\begin{tabular}{ll}
\tableline
\tableline
Wavelength Coverage: & 905--1187\AA \\
\smallskip
Resolving Power: & $\lambda/\Delta\lambda$\,$\approx$\,24,000--30,000 \\
\smallskip
Effective Area: & 20--80 cm$^2$ (beginning of life) \\
\smallskip
Expected Degradation in A$_{eff}$: & $\sim$20\% yr$^{-1}$\\
Science Apertures: & HIRS: (1.25$\arcsec$x20$\arcsec$), MDRS: 
(4$\arcsec$x20$\arcsec$), LWRS:(30$\arcsec$x30$\arcsec$) \\
Point Source Sensitivity: \\
\scriptsize (MDRS aperture, S/N=10, R$>$24,000) \small
& \phantom{0}1 ksec at $F_{1030}$ = 1$\times$10$^{-12}$ erg~cm$^{-2}$~s$^{-1}$~\AA$^{
-1}$ \\
& 10 ksec at $F_{1030}$ = 1$\times$10$^{-13}$ erg~cm$^{-2}$~s$^{-1}$~\AA$^{-1}$ \\
\medskip
& 70 ksec at $F_{1030}$ = 2$\times$10$^{-14}$ erg~cm$^{-2}$~s$^{-1}$~\AA$^{-1}$ \\
\smallskip 
Bright Limit (point source): & $F_\lambda$ = 1$\times$10$^{-10}$ erg~cm$^{-2}$~s$^{-1
}$~\AA$^{-1}$ \\
\smallskip
Dark Limit (approximate): & $F_\lambda$ $\sim$ 3$\times$10$^{-15}$ erg~cm$^{-2}$
~s$^{-1}$~\AA$^{-1}$ \\
\smallskip
Pointing Stability: & 0.5$\arcsec$ in pitch and yaw (FES assisted) \\
\smallskip
Point Spread Function: & 1.5$\arcsec$ (90\% encircled energy) \\
\smallskip
FES Limiting Magnitude: & V\,$\approx$\,15 \\
\smallskip
FES Clear Field of View: & 19$\arcmin\,\times$\,19$\arcmin$ \\
\tableline
\end{tabular}
\end{center}
\end{table}

\begin{figure}[!h]
\includegraphics{sembach1_fig7.ps}
\vspace{2.9in}
\footnotesize 
{\small Figure 7.} FUSE predicted effective area at the beginning of the 
mission.  The individual detector segment coverages are indicated above the
curve (4 segments - 2 SiC and 2 LiF per detector).  The abrupt drops in 
effective area in some locations (e.g., 1082--1085\AA) are due to gaps 
where the individual channels do not overlap completely.
\end{figure}

\section{Launch and Operations} 
NASA will launch FUSE on a Delta\,II-7320 rocket into a 775 km circular, 
25$\deg$ inclination orbit in 1999 from the Cape Canaveral Air Station in 
Florida.  During in-orbit checkout and early operations contact will
be provided through a ground station in Hawaii.
Once operational, the primary FUSE ground station at the 
University of Puerto Rico, Mayaguez will be used for most communications.  
Short duration ($<$13 minutes per orbit) S-band 
communications will occur 6--8 orbits per day, during which time data and 
commands will be sent between the spacecraft and the ground station.  
Communications between the ground station and the satellite control center 
on the Johns Hopkins University  Homewood Campus will occur through an 
ISDN line.
Observation planning, spacecraft instruction commanding, and pipeline 
reduction of scientific data will be performed in the FUSE Operations 
and Science Centers at Johns Hopkins. FUSE is the first mission of its kind 
to be developed and operated within a university setting.

\section{Additional Information}
Technical information about the initial performance results for FUSE
have been given by Wilkinson \et (1998) and Sahnow \et (1998).
Information about observing with FUSE can be found in the {\it FUSE Observer's
Guide} (Oegerle \et 1998) and on the FUSE web site at
{\tt http://fuse.pha.jhu.edu}.  
Guest Investigator questions can be directed to the 
GSFC FUSE Project Scientist, Dr. George Sonneborn, at 
sonneborn@stars.gsfc.nasa.gov.

\acknowledgments
It is a pleasure to thank the many dedicated people who are working so 
hard to make the FUSE mission happen.  I thank Bill Oegerle and 
Ed Murphy for providing an initial 
version of the algorithm used to create Figures 5 and 6, and acknowledge
useful conversations about H$_2$ and HD with Eric Burgh and Steve McCandliss.

\end{document}